\documentstyle[12pt]{article}
\begin{document}
\pagestyle{empty}
\vspace* {13mm}
\baselineskip = 24pt

%
\begin{center}
{\bf PARASTATISTICS AS  EXAMPLES OF THE EXTENDED HALDANE STATISTICS}
\\[8.0mm]

S.Meljanac$^{+}$  \\
 Rudjer Boskovic  Institute, Bijenicka c.54, 10001 Zagreb,
Croatia\\[3mm] 
$^{+}$ E-mail: meljanac@thphys.irb.hr\\[7mm]

M.Milekovic $^{++}$\\
 Prirodoslovno-Matematicki Fakultet, Zavod za teorijsku fiziku, 
 Bijenicka c.32,
\\ 10000 Zagreb, Croatia \\[3mm] 
$^{++}$ E-mail: marijan@phy.hr
\end{center}
\setcounter{page}{1}
\bigskip

{\bf Abstract}

\vskip 0.2cm
%
We show that for every algebra of creation and annihilation operators with a  
Fock-like representation, one can define extended Haldane statistical 
parameters in a unique way. 
Specially for parastatistics, we calculate  extended Haldane parameters 
and discuss the corresponding partition functions.
               
\begin{center}
PACS numbers : 05.30.-d , 71.10.+x 
\end{center}
%
\newpage

\pagestyle{plain}
\def\leer{\vspace{5mm}}
\baselineskip=24pt
\setcounter{equation}{0}
${\bf Introduction}$.- There are two distinct approaches to generalized statistics: (i) the first
 approach obeys some symmetry 
principle (permutation group, braid group, quantum groups and algebras, etc.) 
(ii) the second approach is defined by counting the number of independent multiparticle 
quantum states .\\
The first approach can be characterized by an operator algebra of  creation and 
annihilation operators with a Fock-like representation (for example, parastatistics$^ {1,2}$, 
infinite quon statistics $^{3}$, anyons$^{4}$, etc.).Parastatistics$^ {1,2}$  were the first
 consistent generalization 
of Bose and Fermi statistics in any spacetime dimension. The N-(para)particle 
states can occur in nontrivial representations  of the permutation group 
$S_{N}$, i.e. not only in totally symmetric  (for bosons) or totally antisymmetric 
 (for fermions).\\
 The second approach  is characterized by some Hilbert space of quantum states, 
 generally without a direct connection with creation and annihilation operators acting on
  Fock-like space. This class includes the  recently   suggested Haldane generalization of the 
  Pauli exclusion 
  principle, interpolating between Bose and Fermi statistics . 
  The Haldane statistics ${\it g}$ of a particle is defined by $^{5}$ 
\begin{equation}
g=\frac{d_n -d_{n + \Delta n}}{\Delta n },  
\end{equation}
where ${\it n}$ is the number of particles and ${\it d_n}$ is the dimension of the one-particle 
Hilbert space obtained by keeping the quantum numbers of ${\it (n-1)}$ particles fixed. For 
bosons, ${\it g}=0$ and for fermions, the Pauli principle implies  ${\it g}=1$.\\
Although  Haldane statistics were proposed in any spacetime dimension, all examples known
so far (Calogero-Sutherland model$^{6}$ , 1D spinons$^{5}$ with ${\it g}=\frac{1}{2}$ ,
 anyons residing in the lowest Landau level in a strong magnetic field $^{7}$ ) are essentially 1D   
systems. Furthermore, there is no clear connection between Haldane exclusion statistics and the 
operator algebra.In particular, Green's parastatistics $^{1}$  which also generalize the Pauli exclusion principle 
have  not  been described as a type of Haldane statistics. It is natural to ask the question:is it possible 
to define the ${\it extended}$ Haldane statistics interpolating between  para-Bose and para-Fermi statistics as 
Haldane statistics interpolates between the Bose and Fermi statistics.\\
In this Letter we show that  any operator algebra of  creation and 
annihilation operators with a Fock-like representation can be described in terms of the ${\it extended}$
 Haldane statistics parameters. We apply it to the para-Bose and para-Fermi algebras $^{1,2}$ of 
 M oscillators and calculate few lowest extended Haldane parameters for illustration. 
 They describe parastatistics as the original Haldane parameters $g=0 \;(1)$ describe Bose (Fermi) 
 statistics. 
  We also discuss partition functions for free systems and the counting 
 of independent multiparticle states.

${\bf Extended \; Haldane \;statistics \; parameters}$.- 
Let us start with any algebra of M pairs of creation and 
annihilation operators $a^{\dagger}_{i}$,  $a_i $, $i=1,2,..M $
($a^{\dagger}_{i}$ is Hermitian conjugated to $a_i $ 
). The  algebra is defined by a normally ordered expansion $\Gamma$ (generally 
no symmetry principle is assumed)
\begin{equation}
a_ia^{\dagger}_{j}=\Gamma_{ij}(a^{\dagger};a),
\end{equation}
with the number operators $N_i$, i.e., $[N_{i},a_{j}^{\dagger}]=a_{i}^{\dagger}\delta_{ij}$,
$[N_{i},a_{j}]=-a_{i}\delta_{ij}$.We assume$^{8}$ that there 
is a unique vacuum $|0>$ and the corresponding Fock-like representation.
The scalar product  is uniquely defined by $<0|0>=1$ , the vacuum condition 
$ a_i|0> = 0 $, $i=1,2,..M, $ and Eq.(2).A general N-particle state is a linear combination 
of the vectors ($a^{\dagger}_{i_{1}}\cdots a^{\dagger}_{i_{N}}|0>$),
$i_{1},\cdots i_{N}=1,2,...M.$ We consider Fock spaces with no state vectors of negative 
squared norms.Note that 
we do not specify any relation between the creation (or annihilation) operators themselves.
They appear implicitly as norm zero vectors in Fock space. 
For a state ($a^{\dagger}_{i_{1}}\cdots a^{\dagger}_{i_{N}}|0>$) with fixed indices 
$i_{1},\cdots i_{N}$ we write $1^{n_{1}}2^{n_{2}}...M^{n_{M}}$,
where $n_{1},n_{2}...n_{M}$ are eigenvalues of the number operators $ N_1,\cdots N_M $, satisfying 
$\sum_{i=1}^{M} n_{i}=N$. Then there are $\frac{N!}{n_{1}!n_{2}!...n_{M}!}$ (in principle  
different) states obtained by permutations $\pi \in S_{N}$ acting on the state 
$(a^{\dagger}_{i_{1}}\cdots a^{\dagger}_{i_{N}}|0>)$. From these vectors we form a Hermitian matrix 
${\cal A}(i_{1},\cdots i_{N})$ of their scalar products.The number of linearly independent states 
among them is given by $d_{i_{1},\cdots i_{N}}={\it rank}{\cal A}(i_{1},\cdots i_{N})$.\\
The set of   $d_{i_{1},\cdots i_{N}}$ for all possible $i_{1},\cdots i_{N}= 1,2,\cdots M $ and all 
integers N  completely characterizes the 
statistics and the thermodynamic properties of a ${\it free}$ system with the corresponding 
Fock space.Note that the statistics,i.e., the set 
$d_{i_{1},\cdots i_{N}}$ do not uniquely determine the  algebra given by Eq.(2).
Following Haldane's idea $^{5}$ , we define the dimension of the one-particle subspace 
keeping the ${\it (N-1)}$ quantum numbers $i_{1},\cdots i_{N-1}$ inside the N-particle states, 
fixed : 
\begin{equation}
d_{i_{1},\cdots i_{N-1}}^{(1)}=\sum_{j=1}^{M} d_{j,i_{1},\cdots i_{N-1}}.
\end{equation}
For M Bose oscillators, $d_{i_{1},\cdots i_{N}}=1$ , $d_{i_{1},\cdots i_{N-1}}^{(1)}=M$ for any 
choice $i_{1},\cdots i_{N}$. For M Fermi oscillators, $d_{i_{1},\cdots i_{N}}=1$ 
and 
$d_{i_{1},\cdots i_{N-1}}^{(1)}=M-(N-1)$ if  $i_{1},\cdots i_{N}$ are mutually different 
(otherwise $d_{i_{1},\cdots i_{N}}=d_{i_{1},\cdots i_{N-1}}^{(1)}=0$).
We point out that $d_{i_{1},\cdots i_{N}}$ and $d_{i_{1},\cdots i_{N-1}}^{(1)}$ are integers,
 i.e., no 
fractional dimension is allowed by definition.

Recall that Haldane introduced statistics parameter ${\it g}$ through the change of the single-particle 
Hilbert space dimension ${\it d_n}$ ,Eq.(1). In the similar way we define 
 ${\it extended }$ Haldane statistics parameters $g_{i_{1},\cdots i_{N-1};j_1\cdots j_k}$
through the change of available one-particle Fock-subspace dimension $d_{i_{1},\cdots i_{N-1}}^{(1)}$,
Eq.(3),i.e.
\begin{equation}
g_{i_{1},\cdots i_{N-1};j_1\cdots j_k}=\frac{d_{i_{1},\cdots i_{N-1}}^{(1)} - 
d_{i_{1},\cdots i_{N-1};j_1\cdots j_k}^{(1)}}{k} .
\end{equation}
Note that Eq.(4) implies that  extended  Haldane statistics parameters can be any rational number 
(see also discussion after Eq.(15)).
For bosons, $g_{i_{1},\cdots i_{N-1};j_1\cdots j_k}=0$ , 
$\forall i_{1},\cdots i_{N-1},j_1\cdots j_k$ ,
and for fermions, $g_{i_{1},\cdots i_{N-1};j_1\cdots j_k} = 1$ if 
$i_{1},\cdots i_{N-1},j_1\cdots j_k$
 are mutually different. Note that the quantities $g_{i_{1},\cdots i_{N-1},i_N,j}$, obtained 
 by adding only one particle to the N-particle system, cannot be fractional by definition. This 
 indicates that for the original Haldane statistics with fractional ${\it g}$ , 
 (${\it g } =1/p $ , $ p \in \bf N $),
 Eq.(1), there is no  underlying operator algebra of creation and annihilation
  operators with Fock-like representations. It seems that the original Haldane statistics cannot be 
  realized in the above sense, except for free (or weakly interacting) bosons and fermions 
  (see also discussion 
  in Ref.(5)). \\
 The number of all independent  N-particle states distributed over M quantum states described by M 
 independent  oscillators ($i=1,2 \cdots M$) is given by 
\begin{equation}
D(M,N;\Gamma)= \sum_{i_{1},\cdots i_{N}=1}^{M} d_{i_{1},\cdots i_{N}}. 
\end{equation}
Note that $0\leq D(M,N)\leq M^N$ and $D(M,N)$ is always an integer by definition. For M Bose 
oscillators, $D^B(M,N)=\left( \begin{array}{c}
M+N-1\\ N
\end{array} \right) $ and  for M Fermi oscillators, $D^F(M,N)=\left( \begin{array}{c}
M\\ N
\end{array} \right)$.\\
 Wu $^{9}$ suggested a simple   interpolation 
between the Bose and the Fermi counting
\begin{equation} 
D(M,N;g)=\frac{[ M+(N-1)(1-g)! ]}{N! [ M-gN-(1-g) ]!}
\end{equation}
with ${\it g }=0$ corresponding to bosons and ${\it g }=1$ corresponding to fermions. 
Equation (6) was used in Refs.(9,10) to obtain the partition function and the thermodynamic properties 
(in 1D and 2D systems).
Karabali and Nair $^{11}$ started with the counting rule given by  Eq.(6) and, 
using a few additional assumptions, 
derived an operator algebra defined basically by $(\sum_{i=1}^{M}c_ia_i)^{p+1}=0$, 
$p=\frac{1}{g}\in \bf N $.This operator algebra interpolates
between the Bose ($p=\infty $) and the Fermi ($p=1$) algebra. However, the counting rule (5), 
calculated for the Karabali-Nair algebra,is $D(M,N;p)=D^B(M,N)$ if 
$N\leq p$; $D(M,N;p)=0$ if $N > M p$, and $1\leq D(M,N;p) < D^B(M,N)$ if $p < N \leq M p$, which 
is obviously different from Eq.(6).Moreover, the corresponding partition function for the system with 
the free Hamiltonian 
$H_0=\sum_{i=1}^{M} E_i \, N_i $ is ${\cal Z}(M;p)=\prod_{i=1}^{M}\frac{x_i^{p+1}-1}{x_i-1}$ ,
$x_i=e^{-\frac{E_i}{kT}}$, and leads to the   thermodynamic properties different from those 
obtained in Ref.(9).An ad hoc defined counting formula for $D(M,N)$ can lead in a 
plausible (statistical) way to the partition function and the corresponding thermodynamic properties.
However, the underlying operator algebra may not exist.  

${\bf Parastatistics}$.- The para-Bose and para-Fermi operator algebras are defined by 
trilinear relations 
\begin{equation}
[a^{\dagger}_ia_j \pm a_ja^{\dagger}_i,a^{\dagger}_k]=(\frac{2}{p}) \delta_{jk}a^{\dagger}_i ,
\quad \forall i,j,k =1,2,..M . 
\end{equation}
The upper (lower) sign corresponds to the para-Bose (para-Fermi) algebra,and ${\it p}$ is  the order 
of parastatistics.The unique vacuum $|0>$ is assumed with  the  conditions $<0|0>=1$ , $ a_j|0>=0 $, 
$a_ia^{\dagger}_j|0>=\delta_{ij}|0> $ , $i,j=1,...M.$. The corresponding Fock space does not 
contain any state with negative squared norm only if ${\it p}$  is an integer or 
${\it p}=\infty $.$^{1,2}$ Consistency, i.e. the structure of null-states requires 
\begin{equation}
[a^{\dagger}_i,[a^{\dagger}_j,a^{\dagger}_k]_{\pm}]=0 ,\qquad \forall i,j,k .
\end{equation}
The trilinear relations(7) can be presented in the form of Eq.(2) $^{8}$. Using relations (7) 
and 
(8), one finds by 
induction $^{2,8}$
\begin{eqnarray}
a_ia^{\dagger}_{i_1}\cdots a^{\dagger}_{i_N}|0> & = & \sum_{k=1}^{N} \delta_{ii_k} 
\epsilon ^{k-1}\, a^{\dagger}_{i_1}\cdots \hat{a}^{\dagger}_{i_k}\cdots 
 a^{\dagger}_{i_N}|0> \nonumber \\
 &-& (\frac{2}{p})\sum_{k=2}^{N} \delta_{ii_k}\sum_{l=1}^{k-1}\epsilon ^{l}
a^{\dagger}_{i_1}\cdots \hat{a}^{\dagger}_{i_l}\cdots a^{\dagger}_{i_{k-1}}
a^{\dagger}_{i_l} a^{\dagger}_{i_{k+1}}\cdots a^{\dagger}_{i_N}|0> ,
\end{eqnarray}

where $\epsilon =\mp $,the upper (lower) sign is for parabosons (parafermions) and 
the sign " $\wedge $ " denotes omission of the corresponding operator. The matrices 
${\cal A}(i_{1},\cdots i_{N})$ can be calculated recursively using  Eq.(9). Since Eqs.(7)-(9) are 
invariant under any choice of indices $(1,2,\cdots M)$, the matrix ${\cal A}(i_{1},\cdots i_{N})$ 
 does not depend on any particular choice of 
$(i_{1},\cdots i_{N})$, but it depends only on the corresponding partition $\lambda $ of N ( $N=\sum_{a=1}^{M} 
\lambda_a = |\lambda|$, $\lambda_1 \geq \lambda_2 \geq ...\geq \lambda_M \geq 0$), i.e.,the  
Young tableau of the permutation group $S_N$. Hence, we write ${\cal A}(i_{1},\cdots i_{N})\equiv
{\cal A}_{\lambda}$.\\
If the indices $(i_{1},\cdots i_{N})$ are mutually different, the corresponding partition 
is denoted by $1^N$,with the corresponding $N! \times N!$ 
 ${\it generic}$ matrix ${\cal A}_{1^N}$.All other matrices ${\cal A}_{\lambda}$ ,
 $|\lambda|=N$  are easily obtained from a generic 
 matrix ${\cal A}_{1^N}$ , given by 
\begin{equation} 
{\cal A}_{1^N}=\sum_{\pi \in S_{N}} f(\pi)\, R(\pi) \, .
\end{equation}
Here R is the  right regular representation of the permutation group $S_N$ and 
$f( \pi )$ is a real function  depending on  $\epsilon =\mp 1$ and the order of 
parastatistics ${\it p}$.For ${\it p}$ integer there is no state with negative squared norm.
However, there are null-states, so that $0\leq d_{1^N} \leq N!$.The Hermitian matrix 
${\cal A}_{1^N}$ commutes with every permutation $\sigma $ in the left regular 
representation. Hence, the nondegenerate quotient Fock space splits 
into the sum of irreducible representations (IRREP's) of $S_N$, and we can write 
\begin{equation}
d_{1^N}=\sum_{\mu} n(\mu)\, K_{\mu,1^N} \, \qquad n(\mu)\geq 0,
\end{equation}
where the sum runs over all partitions $\mu $ of N, and $K_{\mu,1^N}$ are Kostka's numbers $^{12}$, i.e., 
the dimension of the irreducible representation (IRREP) $\mu $ of the group $S_N$, 
and $n(\mu)$ is an integer
denoting the multiplicity of IRREP $\mu $,  which can be determined from the spectrum of the matrix 
${\cal A}_{1^N}$.\\
For parastatistics, the function $f( \pi )$, $\pi \in S_N$, in Eq.(11), is given up to $N=4$ by 
\begin{eqnarray}
f(1234)& = & 1,\nonumber \\
f(2134)& = & f(1324)=f(1243)=\epsilon q, \nonumber \\
f(2314)& = & f(3124)=f(2143)=f(1423)=f(1342)=q^2 ,\nonumber \\
f(3214)& = & f(1432)=\epsilon Q ,\nonumber \\
f(2413)& = & f(3142)=f(4123)=f(2341)=\epsilon q^3 ,\nonumber \\
f(3421)& = & f(4231)=f(4312)=\epsilon q (1-2q+2Q) ,\nonumber \\
f(3241)& = & f(4132)=f(4213)=f(2431)= q Q ,\nonumber \\
f(3412)& = & 1-2q^2+2q Q ,\nonumber \\
f(4321)& = & q(2Q-q),
\end{eqnarray}
where $q=1-\frac{2}{p}$ and $Q=q^2+q-1$.\\
Our general result for $d_{\lambda}$, where $\lambda $ is a partition of N, 
$|\lambda |=N \leq 4 $,
obtained from the $24 \times 24 $ matrix ${\cal {A}}_{1^4}$, is given by 
\begin{equation}
d_{\lambda}=rank [{\cal A}_{\lambda} ]=\sum_{\mu;|\mu|=|\lambda |} n(\mu)\, K_{\mu,\lambda} \, ,
\end{equation} 
where, for parabosons, $n(\mu)^{pB}=1$ if $l(\mu)\leq M $ and $ l(\mu)\leq p$ 
(otherwise $n(\mu)^{pB}=0$); for parafermions, $n(\mu)^{pF}=1$ if $l(\mu)\leq M $ and 
 $ l(\mu^T)\leq p$ (otherwise $n(\mu)^{pF}=0$).The $l(\mu)$ and $l(\mu^T)$ are the number of rows of 
 the Young tableaux $\mu $ and $\mu^T $, respectively, and $\mu^T $ denotes the transposed 
 tableau 
 to $\mu $.  Kostka's number $K_{\mu,\lambda}$  is the filling number of an 
 IRREP $\mu $ with independent states arranged according to the partition $\lambda$, 
 $|\mu|=|\lambda |$.  Hence, from all allowed equivalent IRREP's $\mu $, only one appears in the 
 decomposition (13). 
 It is proved$^{13}$ that the pattern for the multiplicities $n(\mu )$ is valid for 
 any M,N,$\lambda$ ,p .
 Specially,for $M=2$ , 
 $\forall \lambda , N, p$,  and for 
 $p=2$ , $\forall \lambda $ , N, M  we reproduce the results of Refs.(14,15). 
  Generally, the thermodynamic properties of a free system are completely defined by 
 the set of all $d_{\lambda}\neq 0$, or by the set of all $n(\mu)$.\\
Let us write the one-particle dimension $d_{i_{1},\cdots i_{N-1}}^{(1)}
 \equiv d_{\lambda}^{(1)}$ given by Eq.(3).
 If the partition $\lambda $ of $(N-1)$ is given by 
 $\lambda_1 \geq \lambda_2 \geq ...\geq \lambda_j > 0$ , $(j\leq M)$, where 
$\lambda_1, \cdots \lambda_j $ are multiplicities of identical quantum numbers among 
$i_{1},\cdots i_{N-1}$, then 
\begin{equation}
d_{\lambda}^{(1)} =  d_{(\lambda_1+1,\lambda_2,\cdots \lambda_j)} + d_{(\lambda_1,\lambda_2+1,\cdots 
\lambda_j)}+ \cdots 
 +  d_{(\lambda_1,\lambda_2,\cdots \lambda_j+1)} + 
( M-j )d_{(\lambda_1,\lambda_2,\cdots \lambda_j,1)}.
\end{equation}  
For example, $d_{0}^{(1)}=M$, $d_{1}^{(1)} =  d_2 + ( M-1 ) d_{(1,1)}$, 
$d_{2}^{(1)} =   d_3 + ( M-1 ) d_{(2,1)}$, \\
$d_{(1,1)}^{(1)} =  2 d_{(2,1)} + (M-2) d_{(1,1,1)}$,
 etc.
Then, the parameters of the extended Haldane statistics  [Eq.(4)] are 
\begin{equation}
g_{\lambda \rightarrow \mu}=\frac{d_{\lambda}^{(1)} - d_{\mu}^{(1)}}{|\mu|-|\lambda|} ,
\qquad \lambda \subset \mu .
\end{equation} 
If $ d_{\mu}^{(1)}=0$, the transition $\lambda \rightarrow \mu $ , $\lambda \subset \mu $, is 
forbidden. A natural definition of exclusion statistics is as follows:  for  every 
 $\lambda $ 
there exists 
$\mu $ , $\lambda \subset \mu $ and $l(\lambda)=l(\mu)$  
 such that $d_{\mu}=0$ . Para-Fermi statistics of order ${\it p} \in \bf N $ 
are  examples of exclusion statistics since at most ${\it p}$ particles can occupy a given 
quantum state, $a_i^{p+1}=0$. For a single oscillator $a^{p+1}=0$, the extended statistical 
parameters are  $ g_{i\rightarrow j}= 0 $ for  $j\leq p$ and 
$g_{i\rightarrow p+1}=\frac{1}{p+1-i}$ .
Para-Bose statistics are not  examples of 
exclusion statistics (except for $p=\infty $).

We present  the extended Haldane parameters $g_{\lambda \rightarrow \mu}$ [Eq.(15)] for some special cases.\\
For parabosons:
$$
\begin{tabular}{|c|c|} \hline
 g & 0 $\rightarrow$ 1 \\ \hline \hline
p=1 & 0 \\ \hline
p$\geq$ 2 & -(M-1) \\ \hline \hline
\end{tabular}
$$
$$
\begin{tabular}{|c|c|c|} \hline
g & 1 $\rightarrow$ 2 & 1 $\rightarrow 1^2$ \\ \hline \hline
p=1 & 0 & 0 \\ \hline
p=2 & 0 & -(M-1) \\ \hline
p$\geq$ 3 & 0 & -(2M-3)\\ \hline \hline
\end{tabular}
$$
For parafermions: 
$$
\begin{tabular}{|c|c|} \hline
 g &  0 $\rightarrow$ 1 \\ \hline \hline
p=1 & 1 \\ \hline
p$\geq$ 2 & -(M+1) \\ \hline \hline
\end{tabular}
$$
$$
\begin{tabular}{|c|c|c|} \hline
g & 1 $\rightarrow $ 2 & 1 $\rightarrow 1^2$ \\ \hline \hline
p=1 & M-1 & 1 \\ \hline
p=2 & M & -(M-3) \\ \hline
p$\geq$ 3 & 0 & -(2M-3)\\ \hline \hline
\end{tabular}
$$
Finally, let us write the partition functions for free parastatistical systems with the Hamiltonian 
$H_0=\sum_{i=1}^{M} E_i \, N_i$, where $N_i$ is the number operator for the $i^{th}$ quantum 
state and $E_i$ is the corresponding one-particle energy. Then the partition function 
${\cal Z}_{N}$ for N-particle states is 
\begin{equation}
{\cal Z}_{N}(x_1,...x_M; p,\epsilon)= \sum_{\lambda;|\lambda|=N} d_{\lambda}(p,\epsilon)
m_{\lambda}(x_1,...x_M) \, ,\qquad x_i=e^{-\frac{E_i}{kT}},
\end{equation} 
where $d_{\lambda}$ is given by Eq.(13) and $m_{\lambda}(x_1,...x_M)$ denotes the monomial $S_M$-symmetric 
function$^{12}$ corresponding to the partition $\lambda$ of N , namely, 
$m_{\lambda}(x_1,...x_M)= \\= \sum_{\pi \in S_M} x_1^{\lambda_{1}}\cdots x_M^{\lambda_{M}}$,
where the sum is over all distinct permutations of $(\lambda_{1},\cdots \lambda_{M})$.\\
Using the  results (13) we can write
\begin{equation}
{\cal Z}_{N}(x_1,...x_M; p,\epsilon)=\sum_{\mu;|\mu|=N}S_{\mu}(x_1,...x_M),
\end{equation} 
where $S_{\mu}(x_1,...x_M)$ are the Schur $S_M$-invariant functions $^{12}$, 
corresponding to the IRREP $\mu$ of $S_M$. The sum in Eq.(17) is restricted to 
$l(\mu )\leq M$ 
and $l(\mu )\leq p$ for parabosons, i.e. to  $l(\mu )\leq M$ and $l(\mu^T )\leq p$ 
for parafermions. Note also that the counting formula (5) is $D(M,N;p,\epsilon)=
{\cal Z}_{N}(\underbrace {1,1,...1}_{M};p,\epsilon)$. 
If $N\leq p$, ${\cal Z}_{N}^{pB}={\cal Z}_{N}^{pF}$ and if $N > p$, 
${\cal Z}_{N}^{pB}>{\cal Z}_{N}^{pF}$.\\
The partition function for M para-Bose oscillators with $p\geq M$ is simple (and generalizes 
the result for $M=2$ of Ref.(15)):
\begin{equation} 
{\cal Z}^{pB}(x_1,...x_M;p\geq M)=\sum_{\lambda}
S_{\lambda}(x_1,...x_M)=\prod_{i=1}^{M}\frac{1}{(1-x_i)}
\prod_{i<j}^{M}\frac{1}{(1-x_ix_j)}.
\end{equation}
The general results for any $S_M$-invariant algebra (2) are \\${\cal Z}_{N}(x_1,...x_M;\Gamma)=
\sum_{\mu} n(\mu) S_{\mu}(x_1,...x_M)$, where $ n(\mu)$ is the multiplicity of 
IRREP $\mu $ in the decomposition (11). The thermodynamic properties of such 
systems will be treated separately.

 ${\it Note \, added}$. In  recent preprints $^{16,17}$,which partly overlap with our paper, the authors 
 follow the first quantized approach to parastatistics.In this Letter we follow  Green's 
 second quantized approach.

${\bf Acknowledgements}$.- We thank D. Svrtan and S.Brant for useful discussions.    

\newpage
\baselineskip=24pt
{\bf References}
\begin{description}

\item{1.}
H.S.Green, Phys.Rev.{\bf 90} (1953) 170;O.W.Greenberg and A.M.L.Messiah, 
 \\Phys.Rev.B {\bf 138} (1965) 1155;J.Math.Phys. {\bf 6} (1965) 500;
Y.Ohnuki and \\S.Kamefuchi, {\it Quantum Field Theory and Parastatistics} 
( University of Tokio Press, Tokio, Springer, Berlin, 1982).
\item{2.}
A.B.Govorkov, Theor.Math.Phys.{\bf 98} (1994) 107.
\item{3.}
O.W.Greenberg,  Phys. Rev.D {\bf 43} (1991) 4111; R.N.Mohapatra, 
Phys.Lett.B, {\bf 242} (1990) 407 ; 
S.Meljanac and A.Perica, Mod.Phys.Lett.A {\bf 9} (1994) 3293.
\item{4.}
J.M.Leinaas and J.Myrheim,  Nuovo Cim. {\bf 37} (1977) 1 ;
F.Wilczek, Phys.Rev.Lett. {\bf 48} (1984) 1144.
\item{5.}
F.D.M.Haldane, Phys.Rev.Lett. {\bf 67} (1991) 937.
\item{6.}
S.Isakov, Int.J.Mod.Phys.A. {\bf 9} (1994) 2563;
 Mod.Phys.Lett.B {\bf 8} (1994) 319;
M.V.N.Murthy and R.Shankar, Phys.Rev.Lett. {\bf 73} (1994) 331;see
 also Comment by A.Dasnieres de Veigy and S.Ouvry,
Phys.Rev.Lett. {\bf 75} (1994) 331.
\item{7.}
A.Dasnieres de Veigy and S.Ouvry, Phys.Rev.Lett. {\bf 72} (1994) 600;
 Mod.Phys.Lett.A {\bf 10} (1995) 1.
\item{8.}
S.Meljanac and M.Milekovic , Int.J.Mod.Phys.A. {\bf 11} (1996) 1391.
\item{9.}
Y.S.Wu, Phys.Rev.Lett. {\bf 73} (1994) 922.
\item{10.}
C.Nayak and F.Wilczek, Phys.Rev.Lett. {\bf 73} (1994) 2740; S.Isakov et al., {\it Thermodynamics 
for Fractional Statistics }, (preprint cond-mat/9601108).
\item{11.}
D.Karabali and V.P.Nair, Nucl.Phys.B {\bf 438 [FS]} (1995) 551.
\item{12.}
I.G.Macdonald, {\it Symmetric Functions and Hall Polynomials} (Claredon, 
Oxford, 1979).
\item{13.}
S.Meljanac,M.Stojic and D.Svrtan, {\it Partition functions for general 
multilevel systems},(preprint 
hep-th/9605064).
\item{14.}
P.Suranyi, Phys.Rev.Lett. {\bf 65} (1990) 2329.
\item{15.}
A.Bhattacharyya,F.Mansouri,C.Vaz and L.C.R.Wijewardhana, 
Phys.Lett.B, {\bf 224} (1989) 384;Mod.Phys.Lett.A {\bf 12} (1989) 1121 .
\item{16.}
S.Chaturvedi, {\it Canonical partition functions for parastatistical system 
of any order}, ( preprint hep-th/9509150).
\item{17.}
A.Polychronakos, {\it Path integrals and parastatistics},(preprint hep-th/9603179).

\end{description}

\end{document}